# Characterization of Visible Range Gain in Praseodymium Doped Fiber Amplifier

**Nasrin Sultana[1], Md. Abubakar Siddik[2]**

[1,2]Department of Electronics and Communication Technology, Hajee Mohammad Danesh Science and Technology University, Dinajpur-5200, Bangladesh

Corresponding author: Nasrin Sultana

**ABSTRACT**

In the optical (380-700 nm) region, a simulation study was conducted to assess the low-signal gain, power conversion efficiency (PCE), and optical output power of a praseodymium-doped fiber optic amplifier (PDFA). The PDFA performance was assessed using the optimized Pr3+ fiber length, Pr3+ ion concentration, and pump power. Additionally, the effects of input signal voltage and amplifier gain on amplified spontaneous emission (ASE) were investigated. A lower peak signal of roughly 1 dB at 635 nm was obtained with a shorter 5 m Pr3 + doped fiber and a higher pump power of 300 mW. The impact of wavelength and pump power variations on the amplifier's optical output power's spontaneous enhanced emission (EEM) is examined. Finally, the ionic interaction (general-purpose conversion effect) in the low-power amplifier signal is analyzed by taking into account various values of the upconversion factor.

**INTRODUCTION**

Ultrafast optics has long been a popular area of research. Over the past ten years, ultrafast-mode fiber lasers have found applications in industrial and medical settings, as well as optical imaging and micromachining [1,2]. Furthermore, it is widely used in the domains of supercontinuum spectrum generation, terahertz generation, and fiber frequency tuning [3-6]. With its short pulse width and high energy, distributed soliton (DS) lasers are among the most attractive types of fiber lasers. DS lasers are capable of producing single-mode beams with up to 31 nJ of power per oscillator, the highest femtosecond pulse energies yet achieved by standard beam lasers [7].

Fiber amplifiers are a promising device for the direct amplification of optical signals in fiber optic transmission systems. Chun Jiang and Li Jin created a theoretical model in 2009 to explain the gain variation along the fiber's length [8]. AV Smith et al. [9] provided a numerical demonstration of the population inversion saturation and stimulated thermal Rayleigh amplification associated with laser amplification of a large-mode fiber amplifier. Chen Peiying et al. (2012) [10] suggested a numerical Euler method solution to the fiber optimizer's frequency equation's initial value problem. Because of the unique properties and possible advantages of active fibers, heavily doped phosphate fibers were used [11].

In the past few years, a great deal of research has been conducted on doped Fber amplifiers using rare earth metal dopants like praseidium, erbium, thuleum, and ytterbium [12–18]. A small doping radius is advised for the erbium-doped fiber amplifier (EDFA) in order to maximize gain and accomplish effective pumping [19]. Er3+, Yb3+, and Pr3+-based synthetic fiber enhancers have been proposed in [20, 21]. The measurement of the amplification efficiency was done at infrared region. The proposed advanced composite fiber amplifier produced a 38 dB raw power variation and less than 3 dB gain variation between 1.53 and 1.565 µm [22]. Aluminum, erbium, zirconium, and yttrium were among the gain media. The thulium-doped fiber amplifier (TDFA) shows a signal gain of approximately 41 dB, 45 dB, and 30 dB at 1.9 µm, 1.95 µm, and 2 µm, respectively. A variety of effective PDFA models operating at approximately 1.3 µm are available in References [23, 24]. By varying the Judd-Ofelt parameters, the effect of the stimulated emission cross-section can be enhanced by taking into account the effects of temperature on Pr3+ ion concentration, ASE interference, and device fabrication components. With an energy conversion efficiency of at least 4.5 µm, the Pr3+ ion-doped broadband mid-infrared chalcogenide fiber amplifier boasts a high gain of 25 dB and 62.8 dB in the 4-5 µm range. 4.3-5.3 µm, in that order [25, 26].

As the previous discussion showed, none of the papers deal with ultrafast lasers with visible range.This work evaluates the performance of the PDFA in the visible range through a comprehensive simulation-based analysis using OptiSystem, a commercial software tool from Optiwave Corporation [27]. Pr3+ doped at the optimal dopant concentration was used to create a high-efficiency optical amplifier model that maximized both the pump power and the fiber length. Therefore, it is thought that by using the suggested PDFA design, developers can maximize the performance of the finished product.

**MATERIALS & METHODS**
The development of doped fiber amplifiers, which can boost signals in the optical domain and supply several optical wavelengths with a high gain. Doped fiber amplifiers can be made with rare-earth dopants, primarily erbium, thulium, praseodymium, and ytterbium [28]. Figure 1 shows the emission spectra of several rare-earth doper fibers.

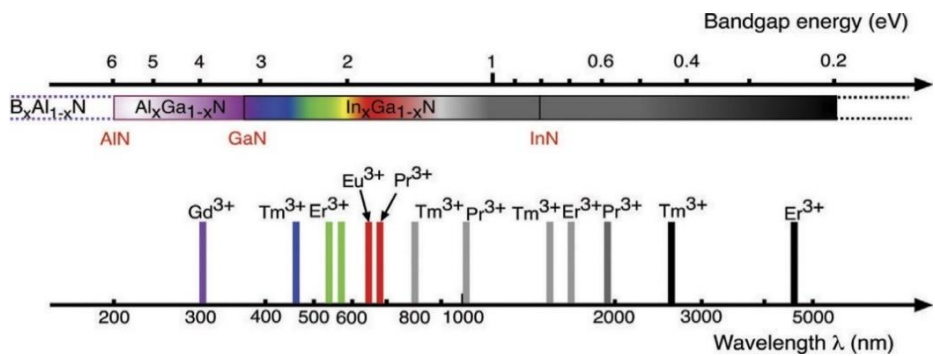

Figure 1. Emission spectra from rare-earth doped fiber

High conversion rates and visible output are offered by fiber-based lasers, using internal frequency shift are gaining popularity. An optical fiber with a large surface area that enables efficient cooling serves as the active medium. Because of its optical characteristics, an optical fiber's transverse mode quality is significant. Because of their red emission, fiber lasers can replace low-energy, high-energy ion or metal vapor lasers in a variety of other applications, such as biological and medical ones. Blue and green have constrained bandwidths. As a result, research focuses on employing diode laser pumps to boost emission in various spectral lines and enhance overall performance.

When a medium absorbs two infrared (IR) photons and responds by releasing one visible photon, this phenomenon is known as lasing by up conversion. Physically, stepwise excitation through an intermediate level populates a highly excited energy level. When a level transitions from an intermediate level to the ground state, lasing may happen. One pump source is needed if the intermediate level's energy is half that of the highly excited level.

The nearly ideal energy-level scheme of the praseodymium ion (Pr3+) allows it to cross the intermediate 1G4 level at 850 nm and travel two photons to reach the highly excited level (Fig. 2). Conversely, the maximum cross section of the ground state to 1G4 transition is approximately 1035 nm. This excitation is inefficient because it only occurs by a "wing" transition when pumped at 850 nm (step 1). Between 840 and 860 nm is the optimal wavelength for the second transition (step 2) from 1G4 to 3P0. This mismatch in excitation wavelengths can be fixed in one of two ways: either employ a second pump wavelength of 630 nm, or convert 850 nm pump photons using the proper co-dopant.

This makes use of Yb3+. Step 3 absorbs four total IR photons, which effectively populates its 2F5/2 level in a nonradiating process. Step 4 involves resonant energy transfer to an alternative Pr3+ ion. As a result, two extremely excited Pr3+ ions are prepared for laser production. This pumping scheme is called an avalanche process because three more photons are absorbed more efficiently after the first photon is absorbed less efficiently. After pumping to the lowest sublevel and completing radiation-free decay, the highly excited 3P0 of Pr3+ is finally ready to emit a competing transition to the level of the ground multiplet to be used as a lasing transition.

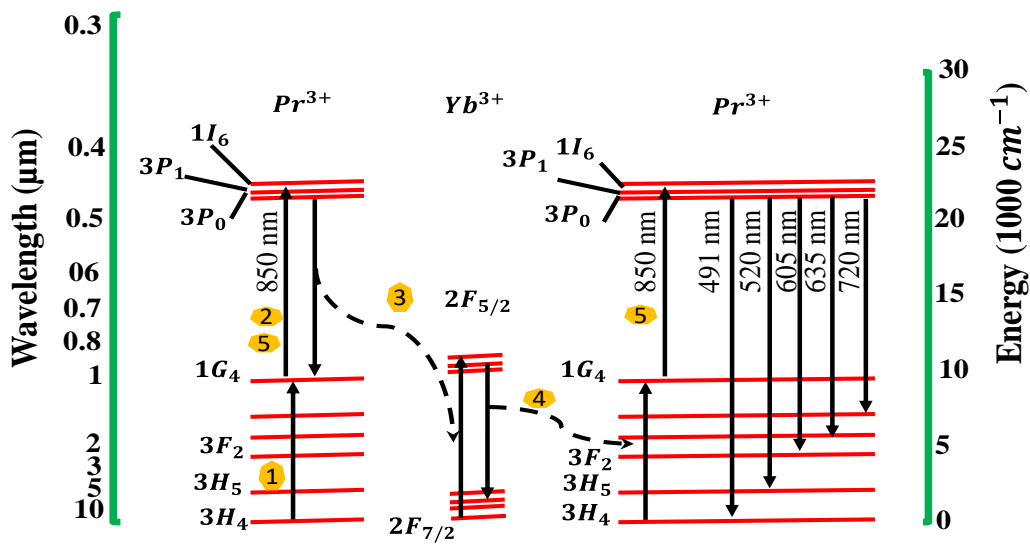

**Figure 2: Upconversion process of praseodymium ion**

The numbers $1G_4$, $3P_0$, $3P_1$, $1I_6$, $3F_2$, $3H_5$, and $3H_4$ represent the primary energy level transitions. The GSA pump and emission are made possible by the transition from 1G4 to 3H4. Similarly, the transitions $1G_4 \rightarrow 3H_5$ and $3H_4 \rightarrow 3F_4$ hold signal emission and ground state absorption (GSA). Furthermore, since the energy variance concerning the 1G4 and 1I6 levels equals the energy variance concerning the 1G4 and 3H5 levels, the up-conversion effect caused by the $1G_4 \rightarrow 1I_6$ and $1G_4 \rightarrow 3H_5$ transitions can lower the carrier concentration of the 1G4 level [29]. The overall density, $n_t$, is determined by the carrier concentrations at each level, signified by the numbers $n_1$, $n_2$, $n_3$, $n_4$, and $n_5$.

$$n_t = n_1 + n_2 + n_3 + n_4 + n_5 \tag{1}$$

The rate equations for the Pr3+ energy level diagram in Fig. 2 can be expressed as follows:

$$\frac{dn_3}{dt} = \gamma_{13}n_1 - \left(\gamma_{35} + \gamma_{34} + \gamma_{32} + \gamma_{31} + \frac{1}{\tau_3} + cn_3\right)n_3 + \frac{B_{43}}{\tau_4}n_4 + \frac{B_{53}}{\tau_5} \quad (2)$$

$$\frac{dn_4}{dt} = \left(\gamma_{35} + \frac{C}{2}n_3\right)n_3 - \frac{n_4}{\tau_4} \quad (3)$$

$$\frac{dn_5}{dt} = \gamma_{35}n_3 - \frac{n_5}{\tau_5} \quad (4)$$

In Eq. (2) to (4), the transition rates $\gamma_{13}, \gamma_{31}, \gamma_{31}, \gamma_{34}$, and $\gamma_{35}$ are given by $\frac{P_p\sigma_{13}np}{A_ch\nu_p}, \frac{P_p\sigma_{31}np}{A_ch\nu_p}, \frac{P_p\sigma_{32}np}{A_ch\nu_p}, \frac{P_p\sigma_{34}np}{A_ch\nu_p}$ and $\frac{P_p\sigma_{35}np}{A_ch\nu_p}$, respectively. Similarly, $B_{53}$ and $B_{43}$ are branching ratios for the $^3P_0 \rightarrow {}^1G_4$ and $^1I_6 \rightarrow {}^1G_4$ transitions.

The transitions between the 3H4 and 1G4 levels, which have carrier densities of n1 and n3, respectively, dictate the PDFA's full small-signal gain. The signal propagation equation can be expressed as follows [29] when the PDFA with a depth of dz travels using PDFA as a gain medium:

$$\frac{dP_s}{dz} = [n_3(\sigma_{32} - \sigma_{34}) - n_1\sigma_{13}]P_s\Gamma_s \quad (5)$$

The excited state absorption (ESA) cross-section at the signal wavelength (σ 34) is not taken into account by OptiSystem in the given expression. Thus, the obtained signal efficeiency is given by [29].

$$G = \frac{P_s(L)}{P_s(0)} = exp\left[L\Gamma_s\{n3\,(\sigma32 - \sigma34) - n1\sigma13\}\right] \quad (6)$$

where PDF length is L.

## SIMULATION SETUP

The proposed PDFA implementation plan is illustrated in Fig. 3. Two optical isolators, a WDM coupler, a laser pump source, and a brief fiber section with Pr3 + ions make up the amplifier. Optical isolation is employed in this configuration to minimize performance-degrading back emissions, stabilize amplifier operation, and stop the amplifier from operating as a laser. A laser with a wavelength of p = 6.3 μm and a pump power of 300 mW as the gain medium PDFA. Short wavelengths are used in indirect pumping to activate the Pr3+ ions in the PDFA growth medium. Usually, the ground area (GSA) is injected using short wavelengths up to 1.03 μm. A brief wavelength of 840 can be employed for indirect excitation.

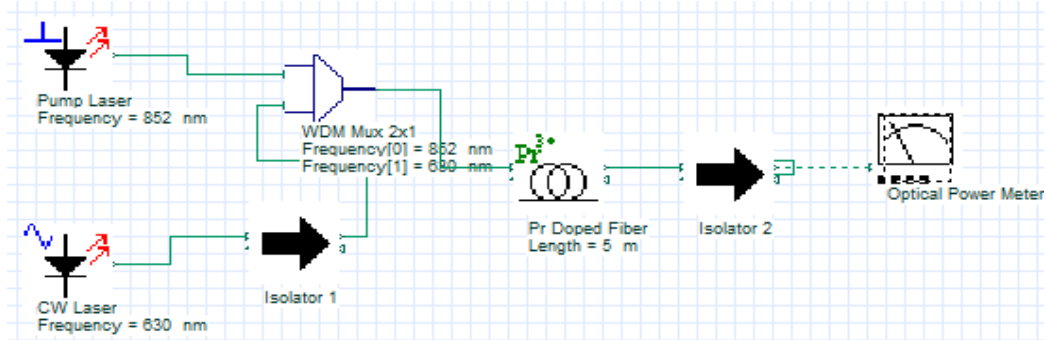

**Figure 3.: Schematic of the proposed setup for PDFA**

## RESULTS & DISCUSSION

Figure 4 considers the optimized MFD, Pr3+ concentration, and PDF length for a signal power of -65 dBm. The signal wavelength is plotted against the pump power and the ASE noise power.

Using a signal wavelength of 850 nm and a pump power of 300 mW, it can be seen that the highest ASE is near -45 dBm. Due to spontaneous emission, more photons are produced during optical amplification as the pump power increases. As a result, the sound power of ASE increases with pump power. Furthermore, at longer wavelengths, the ASE exhibits a significant decreasing trend, which could be explained by the absence of population inversion at these longer wavelengths.

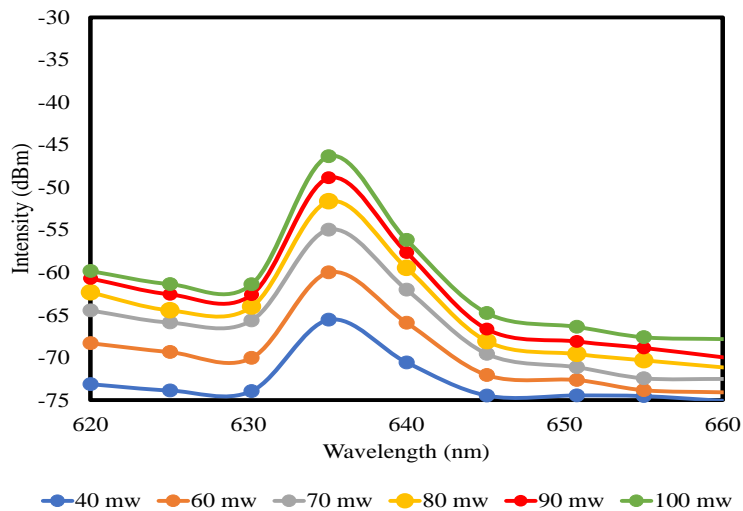

**Figure 4: Signal wavelength plotted against ASE as a function of pump power**

The up-conversion effect, an ion-ion interaction that impacts PDFA performance, is connected to Pr3+ concentration in PDF. Up-conversion usually has a negative effect on PDFA performance when PDF has a high Pr3+ concentration (nt> 50 × 1024 m−3). Even at high pump powers, about 10% of Pr3+ ions in highly doped PDFs stay in the ground state. Pumping efficiency is lost due to a fast up-conversion that occurs when both ions in a pair are pumped.

By varying the length of the praseodymium doped fiber at various pump strength, one can perceive the progression of the gain of the PDFA when the concentration of Pr3+ ions is 50 × 1024 ions m−3, as shown in Fig. 5a. It is clear that when the 5 m PDF and 100 mW pump power are used, the PDFA's maximum gain is about 1 dB. The trend of decreasing gain increases with PDF length and is caused by a decline in population inversion. Thus, 5 m is found to be the optimal length for the PDF that produces the highest gain. Gain versus input signal power plots for various Pr3+ ion concentrations are displayed.

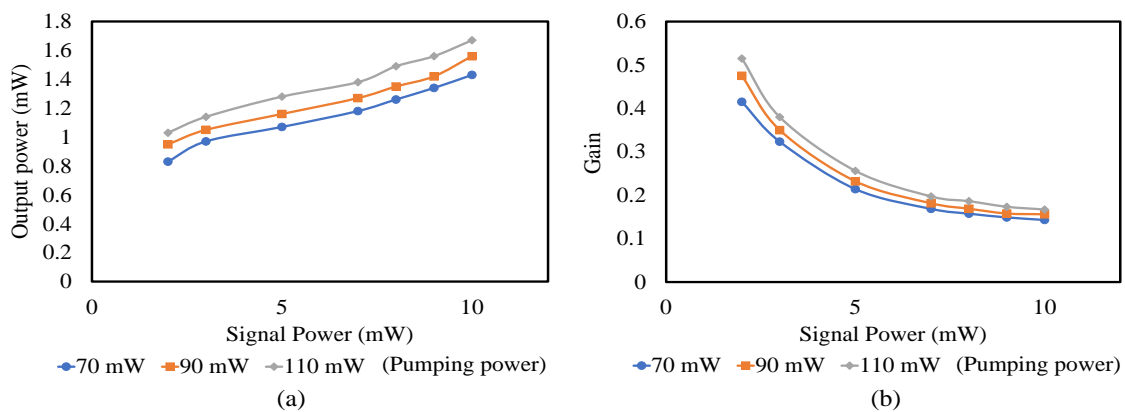

**Figure 5: (a) Signal power *vs*. Output power (b) Signal power vs. gain at different pump power**

To estimate the PCE of the PDFA, draw pump control against output control as a province of input signal energy for a given PDF length. Figure 6a illustrates that the PCE value reaches its maximum of 4% at 8.8 dBm of input signal power and its minimum of 2% at 2.3 dBm of input signal power. Scheming the pump wavelength contrary to the output strength of the input signal allows to examine how a change in the pump wavelength affects the output power of the amplifier.

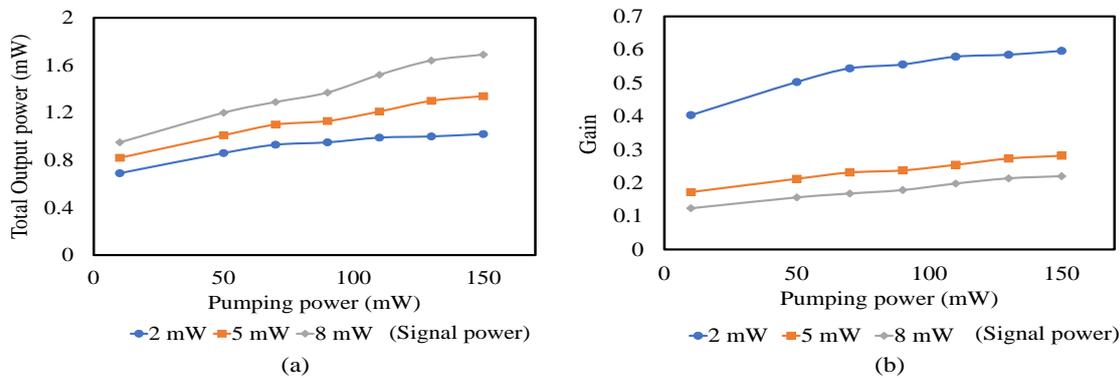

**Figure 6: (a) Pumping power *vs*. Output power (b) Pumping power vs. gain at different signal power**

## CONCLUSION

Opti_System simulation files are used to evaluate the praseodymium doped fiber amplifier's performance and visualization. The findings indicate that an optimized praseodymium doped fiber length of roughly 5 m can achieve a peak gain of about 1 dB for an maximized pumped power of 300 mW, and input signal wavelength of 635 nm. This discovery has important implications for all visible fiber ultrasort pulsed lasers.

*Declaration by Authors*
**Acknowledgement:** None
**Source of Funding:** None
**Conflict of Interest:** The authors declare no conflict of interest.

\*\*\*\*\*\*